\documentclass{elsart5}

 \usepackage{graphicx}

\usepackage{amssymb}
\usepackage{amsmath}

\begin{document}

\begin{frontmatter}

\title{The de Haas-van Alphen effect in Kondo systems with crystalline electric field}

\author{Junya Otsuki\corauthref{cor1}}
\ead{otsuki@cmpt.phys.tohoku.ac.jp}
\corauth[cor1]{}
\author{Hiroaki Kusunose}
\author{Yoshio Kuramoto}
\address{Department of Physics, Tohoku University, Sendai 980-8578}
\received{12 June 2005}
\revised{13 June 2005}
\accepted{14 June 2005}


\begin{abstract}
The effective mass probed by the de Haas-van Alphen oscillations is studied 
for a model Ce system under magnetic field higher than the Kondo energy.  
In the mean-field theory, the mass enhancement per Ce ion in the periodic system is identical with that in the dilute system.  
With decreasing magnetic field,  the effective mass tends to diverge corresponding to a formation of the Kondo ground state. 
The effective mass can be very different between up and down spins depending on the nature of the $4f$ wave functions.
\end{abstract}


\end{frontmatter}


The de Haas-van Alphen (dHvA) effect is a powerful measure 
to observe the quasi-particle behavior of each spin component separately. 
In dilute Kondo systems, earlier work ~\cite{Miwa,Simpson,Shiba,Fenton} concerned about the spin splitting and the Dingle temperature.
In more recent dHvA experiments on CeB$_6$, which shows a typical Kondo effect, 
it has been found that one of spin components does not contribute to oscillatory magnetization~\cite{Joss,Harrison}.
Furthermore, the cyclotron effective mass of the observed spin component tends to diverge as applied magnetic field decreases~\cite{Endo,Nakamura}. 
Since enhancement of the effective mass in Ce$_x$La$_{1-x}$B$_6$ is proportional to Ce concentration $x$~\cite{Endo,Nakamura,Goodrich},
the origin of the anomalous mass enhancement as well as its spin dependence are ascribed to the single-site effect. 
The purpose of this paper is to clarify the influence of the Kondo effect on the cyclotron effective mass starting with the dilute limit.   The results are compared with those obtained for periodic systems with $SU(N)$ symmetry~\cite{Wasserman}.

The oscillatory part $M_{\rm osc}$ of magnetization per unit volume is given in terms of 
the cross-sectional area $S$ of the Fermi surface, 
the cyclotron frequency 
$\omega_{\rm c}=eH /mc$ 
without many-body effect, and the self-energy $\Sigma_{\sigma}({\rm i}\omega_n)$ of conduction electrons with $\omega_n=(2n+1)\pi T$ the fermion Matsubara frequency
as follows ($\hbar =k_{\rm B}=1$)~\cite{Shiba,Engelsberg}:
\begin{align}
	\frac{M_{\rm osc}}{V} 
	&= \frac{e^{1/2} T}{(2\pi^3 c H)^{1/2}} \sum_{\sigma=\pm} \sum_{l=1}^{\infty}
	 S \left| \frac{\partial^2 S}{\partial k_z^2} \right|^{-1/2} \nonumber 
\end{align}
\begin{align}
	&\times {\rm Re} \left\{ {\rm i} \exp \left[ 2\pi {\rm i}l
	 \left( \frac{Sc}{2\pi eH} -\phi \right)
	  \mp \frac{{\rm i}\pi}{4} \right] \right. \nonumber \\
	&\times \left. \sum_{n=0}^{\infty} \exp \left[ \frac{2\pi {\rm i}l}{\omega_{\rm c}}
	  ({\rm i}\omega_n -\sigma \mu_{\rm B}H -\Sigma_{\sigma}({\rm i}\omega_n)) \right]
	\right\}.
	\label{eq:mag_osc}
\end{align}
The standard Lifshitz-Kosevich formula is recovered by setting $\Sigma_{\sigma}({\rm i}\omega_n)=0$ in eq.~(\ref{eq:mag_osc}).

In the dilute systems ($x\ll 1$) with completely random distribution of impurities, $\Sigma_{\sigma}({\rm i}\omega_n)$ is given by 
$\Sigma_{\sigma}({\rm i}\omega_n) = x t_{\sigma} ({\rm i}\omega_n)$, where $t_{\sigma} ({\rm i}\omega_n)$ denotes the impurity $t$-matrix~\cite{Shiba_commun}.
The $t$-matrix is diagonal in the crystalline-electric-field (CEF) basis
$\alpha$ at the impurity site. 
In terms of the Bloch basis $(k,\sigma)$, which describes the cyclotron motion, the $t$-matrix is given by
\begin{align}
	t_{\sigma}(\omega) = \sum_{\alpha} w_{\sigma, \alpha} t_{\alpha}(\omega),
\end{align}
where the momentum dependence is neglected.
The transformation coefficient $w_{\sigma, \alpha}$ depends strongly on the CEF states.

The component $n=0$ gives a dominant contribution 
near of the Fermi level in eq.~(\ref{eq:mag_osc}), provided
$2\pi^2 T/\hbar \omega_{\rm c} \gg 1$.
We expand $\Sigma_{\sigma}({\rm i}\omega_0)$ as follows:
\begin{align}
	\Sigma_{\sigma}({\rm i}\omega_0) \simeq
	 \Sigma_{\sigma}(0)
	  + {\rm i}\omega_0 \partial \Sigma_{\sigma}(\omega) / \partial \omega |_{\omega=0}.
\end{align}
Consequently, ${\rm Re} \Sigma_{\sigma}(0)$ 
gives a shift of the dHvA frequency,
and ${\rm Im} \Sigma_{\sigma}(0)$ yields 
the relaxation rate or the Dingle temperature.
The cyclotron mass $m$ is replaced by a spin-dependent effective mass 
$m_{{\rm c}\sigma}^*$ defined in terms of $t_{\sigma}(\omega)$ by
\begin{align}
	m_{{\rm c}\sigma}^* /m -1
	 = -x\partial {\rm Re}t_{\sigma}(\omega)/\partial \omega |_{\omega=0}
	 \equiv x \mu_{\sigma}
	\label{eq:mass_enh}
\end{align}
In this paper we restrict our consideration to the effective mass, relegating the spin splitting to another publication. 


It is well-known that the ground state of the Kondo impurity system is a local Fermi liquid, provided the number of internal degrees of freedom matches the number of screening channels.  However the lifetime of conduction electrons, which appears in the dHvA oscillation, can remain finite since it always refers to the Bloch basis.
If the lifetime is short,   
the effective mass as defined by eq.~(\ref{eq:mass_enh}) can take even  a negative value, and is not related to an observable.
In such a case, the impurity specific heat is not determined by $m_{{\rm c}\sigma}^*$, but by the phase shift in the local Fermi liquid.
In order to demonstrate the point in the simplest manner, we take the 
$N$-fold degenerate Anderson model with constant hybridization $V$ and 
infinite Coulomb repulsion $U$, and use the slave boson mean field approximation (SBMFA).   In contrast with the periodic $SU(N)$ model,  however, the Bloch state in our impurity model has only the spin degeneracy. 
The $4f$ electron Green function is given by
\begin{align}
	G_{f,\alpha}(\omega^+) = r^2 [ \omega^+ - (\tilde{\epsilon}_f - h_{\alpha})
	 +{\rm i} \tilde{\Delta}]^{-1},
	\label{eq:Gf}
\end{align}
where 
$r^2$ is the renormalization factor, 
$h_{\alpha}=g_{\alpha} \mu_{\rm B}H$ is the Zeeman shift, 
and we have introduced $\omega^+ = \omega + {\rm i}0$ and $\tilde{\Delta}=r^2 \Delta=r^2 \pi V^2 \rho_{\rm c}$. 
We take a constant density of states $\rho_{\rm c}=1/2D$ with $2D$ being band width.
The impurity $t$-matrix is then given by
$
t_{\alpha}(\omega^+) = |V|^2 G_{f,\alpha}(\omega^+)
$,
and the enhancement factor $\mu_{\alpha}$ of each channel are given by
\begin{align}
	\mu_{\alpha}
	= -\left. \frac{\partial {\rm Re} t_{\alpha}(\omega)}{\partial \omega} \right|_{\omega=0}
	= \frac{\tilde{\Delta}}{\pi \rho_{\rm c}}
	 \frac{(\tilde{\epsilon}_f - h_{\alpha})^2 - \tilde{\Delta}^2}
	  {[(\tilde{\epsilon}_f - h_{\alpha})^2 + \tilde{\Delta}^2]^2}.
\label{eq:slope_imp}
\end{align}
In the high magnetic field regime, $|\tilde{\epsilon}_f - h_{\alpha}| \gg \tilde{\Delta}$, eq.~(\ref{eq:slope_imp}) becomes
\begin{align}
	\mu_{\alpha}
	\simeq \frac{\tilde{\Delta}}{\pi \rho_{\rm c} (\tilde{\epsilon}_f - h_{\alpha})^2}.
\label{eq:slope_imp_high}
\end{align}
In this regime,  the enhancement factor determines the physical effective mass which is relevant to the specific heat.

For large-$N$, renormalized parameters are given in terms of the Kondo temperature $T_{\rm K}$ by $\tilde{\epsilon}_f = T_{\rm K}$ and $\tilde{\Delta} = \pi n_f T_{\rm K} /N$~\cite{Hewson}.
In this limit, eq.~(\ref{eq:slope_imp_high}) agrees with
the corresponding quantity obtained for periodic systems~\cite{Wasserman}. 
Hence under high magnetic field, where 
the Kondo singlet becomes unstable,
there is no difference between periodic and dilute systems 
in  the mass enhancement, except for the concentration factor. 
We remark that eq.~(\ref{eq:slope_imp_high}) applies even in the weak field
for the periodic system~\cite{Wasserman}. 

\begin{figure}[t]
	\includegraphics[width=0.90\linewidth]{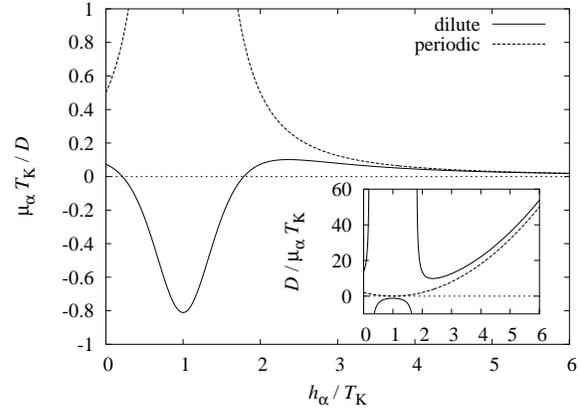}
	\caption{Magnetic field dependence of the enhancement factor $\mu_{\alpha}$ and $\mu_{\alpha}^{-1}$ (inset) of the orbital with $g_{\alpha}>0$ for dilute systems, eq.~(\ref{eq:slope_imp}), and for periodic systems, eq.~(\ref{eq:slope_imp_high}).}
	\label{fig:mass_enh}
\end{figure}
Figure~\ref{fig:mass_enh} shows the enhancement $\mu_{\alpha}$ with $g_{\alpha}>0$ for $N=4$ in the SBMFA. 
Since $D/T_{\rm K}$ is of order $10^4$ for Ce$_x$La$_{1-x}$B$_6$, $\mu_{\alpha}$ 
is of order $10^3$.
In dilute systems, $\mu_{\alpha}$ becomes negative around $h_{\alpha} \sim T_{\rm K}$.  
For periodic systems, on the other hand, $\mu_{\alpha}$ diverges at $h_{\alpha} = T_{\rm K}$ corresponding to the Kondo resonance.

The negative value of $\mu_{\alpha}$ in the dilute systems at $h_{\alpha} \sim T_{\rm K}$ is an outcome of a short lifetime of conduction electrons. 
In addition to the large relaxation, 
low magnetic field with small energy interval of the Landau levels should make the dHvA signals almost invisible.  
Therefore extrapolation of the mass enhancement from high field region  
 leads to a divergence around $h_{\alpha} \sim T_{\rm K}$. 
The inset of Fig.~\ref{fig:mass_enh} shows the inverse effective mass which tends to zero.
On the other hand, in periodic systems at $h_{\alpha} \sim T_{\rm K}$, 
the Fermi surface includes $4f$ electrons. 
Thus the topology should change drastically as the field decreases.

We now apply the above results to Ce$_x$La$_{1-x}$B$_6$.
Spin-dependent enhancement factor $\mu_{\sigma}$ is evaluated by $\mu_{\sigma} = \sum_{\alpha} w_{\sigma, \alpha} \mu_{\alpha}$. 
Hence the character of each signal is dominated by the orbital having the largest value of $\mu_{\alpha}$, i.e., the most stable one in the magnetic field. 
According to a simple calculation with the $\Gamma_8$ CEF wave function, the lowest orbital under $H//[001]$ includes up and down spins in the proportion of 16/21 to 5/21.
Consequently, the conduction band with up spin is more difficult to observe due to the heavier mass and the large relaxation. 
On the other hand, the down spin is easier to observe, and $\mu_{\sigma}$ should behave like Fig.~\ref{fig:mass_enh} against magnetic field. 

We acknowledge valuable discussion with Prof. H. Shiba and Prof. H. Aoki.

\end{document}